\documentclass[12pt]{iopart}   
\usepackage{graphicx}
\usepackage{epsfig}
\usepackage{amssymb}
\begin{document}

\newcommand{\be}{\begin{equation}}
\newcommand{\ee}{\end{equation}}
\newcommand{\beqn}{\begin{eqnarray}} 
\newcommand{\eeqn}{\end{eqnarray}}

\title[Persistence discontinuity in disordered contact processes]{Persistence discontinuity in disordered contact processes with long-range interactions}

\author{R\'obert Juh\'asz}
\address{Institute for Solid
State Physics and Optics, Wigner Research Centre for Physics, H-1525 Budapest,
P.O. Box 49, Hungary}
\ead{juhasz.robert@wigner.hu}

\begin{abstract}
We study the local persistence probability during non-stationary time evolutions in disordered contact processes with long-range interactions by a combination of the strong-disorder renormalization group (SDRG) method, a phenomenological theory of rare regions, and numerical simulations. 
We find that, for interactions decaying as an inverse power of the distance, the persistence probability tends to a non-zero limit not only in the inactive phase but also in the critical point. Thus, unlike in the contact process with short-range interactions, the persistence in the limit $t\to\infty$ is a discontinuous function of the control parameter. 
For stretched exponentially decaying interactions, the limiting value of the persistence is found to remain continuous, similar to the model with short-range interactions.  
\end{abstract}

\maketitle

\section{Introduction}

Local persistence, which is the probability that a local field does cross a given level up to time $t$, attracted a lot of attention \cite{bms,majumdar,redner}. In systems with many degrees of freedom, the time-dependence of persistence during non-stationary time evolutions is a challenging problem and exact results are scarce \cite{derrida}. 
An important class of reaction-diffusion systems in which much numerical effort has been devoted to the study of this question \cite{hk,am,menon,fuchs,grassberger} is the directed percolation (DP) universality class \cite{md,hhl,odor},  
a simple representative of which is the contact process \cite{cp,liggett,md}. Here, sites of a lattice are either empty or occupied by a particle, which can spontaneously annihilate or create another particle on a neighboring lattice site. The persistence $P(t)$ in this model is defined as the probability that an initially empty site remains empty till time $t$. It was found that, in the inactive phase, $P(t)$ tends to a non-zero limit as $t\to\infty$, while, in the active phase, it tends to zero exponentially \cite{hk}. In the critical point, it vanishes algebraically, $P(t)\sim t^{-\Theta}$, where the persistence exponent is universal for several models in the DP class for dimensions $d\le 4$ \cite{hk,am,menon}, for exceptions, see \cite{matte,saif}, otherwise it is model-dependent \cite{fuchs,grassberger}.  

Recently, the effect of quenched disorder on the time-dependence of persistence has been studied in the contact process by means of the strong-disorder renormalization group (SDRG) method \cite{im,hiv} in dimensions $d=1,2$, and $3$ \cite{jk2020}. The average persistence was found to tend to zero in the critical point ultra-slowly  as $P(t)\sim (\ln t)^{-\overline{\Theta}}$, where the generalized exponent $\overline{\Theta}$ is independent of the form of disorder and depends only on the dimension. In $d=1$, it was shown furthermore that the distribution of sample-dependent local persistences is characterized at late times by a limit distribution of effective persistence exponents. According to a phenomenological theory of rare regions, the average persistence in the active phase was found to decay as $P(t)\sim \exp[-const\cdot (\ln t)^d]$, which simplifies to a power law with non-universal exponents in $d=1$. Such type of anomalous decay has also been observed in the active phase of a similar, one-dimensional model with quenched disorder \cite{bg}.

The contact process can be interpreted as a simple model of epidemic spreading, in which empty (occupied) sites represent healthy (infected) individuals. In this context, the persistence $P(t)$ has a quite natural meaning: it is the probability that an initially healthy individual is not infected until time $t$. 

In this work, we go further in exploring the behavior of persistence in disordered contact processes and instead of a nearest-neighbor interaction we consider long-range interactions of two types. A widely studied form is when the strength of interaction, the creation rate in the present case, decays algebraically with the distance as $\lambda(l)\sim l^{-\alpha}$ \cite{mollison,janssen,fisher_me,sak,bloete,picco,parisi,howard,linder,grassberger2,ginelli,adamek,hinrichsen}. 
For the disordered contact process, even a more rapidly decaying interaction of stretched exponential form, $\lambda(l)\sim e^{-const\cdot l^{-a}}$, is able to alter the critical behavior of the short-range model, as it was shown in Ref. \cite{stretched}. Therefore we will consider this form of long-range interactions, as well. Both types of models can be approached by the SDRG method; in the first case, the critical behavior is controlled by a finite-disorder fixed point of the SDRG transformation \cite{jki,qlrcp,kji_3d}, while in the second case by an infinite-disorder fixed point \cite{stretched}.    

We also apply a phenomenological theory of rare regions to infer the time-dependence of persistence in the active phase and confront the results with Monte Carlo simulations. 
For algebraically decaying interactions, we find that, unlike in other variants of the contact process, the average persistence in the critical point does not vanish but tends to a positive constant as $t\to\infty$. The persistence in this limit is thus a discontinuous function of the control parameter. 
For stretched exponential interactions, however, the average persistence tends to zero ultra-slowly in the critical point, the limiting value being a continuous function of the control parameter. 

The rest of the paper is organized as follows. The models and the persistence are defined in section \ref{model}. The SDRG approach of the disordered contact process is reviewed in section \ref{sdrg}, and applied to calculate the persistence for power-law interactions in section \ref{sec:pl} and for stretched-exponential interactions in section \ref{sec:se}. A simple phenomenological theory of rare-region effects in the active phase is discussed in section \ref{phenomenology}. 
In section \ref{numerical}, results of numerical simulations are presented. Finally, results are summarized and discussed in section \ref{discussion}.

\section{Contact processes and persistence}
\label{model}

In the contact process \cite{cp,liggett,md}, the state of the system is specified by a set of binary variables $n_i=0,1$ attached to the sites of a lattice, which is chosen to be one-dimensional in this work. The contact process is a continuous-time Markov process with two kinds of transitions. Occupied sites ($n_i=1$) become spontaneously empty ($n_i=0$) with a rate $\mu_i$, and can create a particle at another site $j$ with a rate $\lambda_{ij}$, provided that site was empty.  We assume that the annihilation rates $\mu_i$ are $O(1)$, independent, identically distributed quenched random variables. 
The creation rates depend only on the distance $l=|i-j|$, and we consider two different forms of $\lambda(l)$. 
One of them is a power law: 
\be 
\lambda(l)=\lambda_0l^{-\alpha},
\label{pl}
\ee   
where the exponent is restricted to $\alpha>1$, so that the total creation rate 
$\sum_{l=1}^{\infty}\lambda(l)$ is finite. 
We shall refer to this case as the power law (PL) model. 
In the other case, $\lambda(l)$ decreases according to a stretched exponential function:
\be 
\lambda(l)=\lambda_0e^{-(l/l_0)^{a}}.
\label{se}
\ee   
Here $\lambda_0$ and $l_0$ are positive constants. 
For $a>1/2$, the critical behavior of this model is the same as that of the disordered contact process with short-range (SR) interactions, whereas, for $0<a<1/2$, the critical exponents vary continuously with $a$ \cite{stretched}. Therefore, concerning the critical point, we restrict ourselves to study the regime $0<a<1/2$, while, in the inactive phase, the approach is valid in a broader range, $0<a<1$. We call this model the stretched exponential (SE) model.

In both cases, the parameter $\lambda_0$ serves as a control parameter of the phase transition. For $\lambda_0<\lambda_c$, the only steady state is the state with all sites empty, while, for $\lambda_0>\lambda_c$, there is also a non-trivial steady state with a non-zero fraction of occupied sites.

We define the local persistence $P_0(t)$ at site $0$ of a fixed realization of disorder as follows. The system starts to evolve from the state in which all but site $0$ are occupied, and $P_0(t)$ is the probability that site $0$ remains empty until time $t$. The average over random samples will be denoted by $\overline{P_0(t)}$.  

Specially, in a one-dimensional system with nearest-neighbor interactions, as long as site $0$ is persistent, there is no interaction between the parts of the system composed of sites $n>0$ and $n<0$. As a consequence, $P_0(t)$ is a product of persistence probabilities $P_{0+}(t)$ and $P_{0-}(t)$ at site $0$ in the subsystems $n\ge 0$ and $n\le 0$, respectively: 
\be 
P_0(t)=P_{0+}(t)P_{0-}(t).    
\label{decomp}
\ee
In this case, it is therefore sufficient to consider the persistence at the first site of a semi-infinite chain. 
In the presence of long-range interactions, the relationship in Eq. (\ref{decomp}) is not exactly valid. However, the SDRG approach of both PL and SE models relies on that the relevant interactions at any stage of the renormalization procedure are those between the nearest-neighbor effective degrees of freedom \cite{jki,qlrcp,stretched}.  
As a consequence, within the SDRG approach, the relationship in Eq. (\ref{decomp}) holds for the PL and SE models, as well. 

There is a useful representation of $P_0(t)$ as a return probability. Let us consider a fixed random sample in which the annihilation rate at site $0$ is set to zero, $\mu_0=0$, and consider the time evolution of the system from the initial state in which only site $0$ is occupied. The return probability to the initial state at time $t$ is exactly equal to $P_0(t)$ \cite{jk2020}.   

\section{The strong-disorder renormalization group method}
\label{sdrg}

The SDRG method is a sequential, real-space renormalization group method \cite{mdh,fisher,im}, which was first applied to the disordered contact process in Ref. \cite{hiv}.  By this procedure, the quickly relaxing degrees of freedom are eliminated one by one, resulting thereby the gradual decrease of the rate scale $\Omega$, which is set by the largest transition rate, $\Omega=\max\{\mu_i,\lambda_{ij}\}$.
There are two kinds of local renormalization steps. If the largest rate is  a creation rate, $\Omega=\lambda_{ij}$, then the variables $n_i$ and $n_j$ are replaced by a single binary variable $n_{ij}$, which has an effective annihilation rate: 
\be 
\tilde\mu_{ij}=2\frac{\mu_i\mu_j}{\Omega}. 
\label{mu}
\ee
If the largest rate is an annihilation rate, $\Omega=\mu_i$, site $i$ is deleted, and interactions between the remaining sites with effective creation rates 
\be 
\tilde\lambda_{jk}=\frac{\lambda_{ji}\lambda_{ik}}{\Omega}. 
\label{lambda}
\ee
are generated.
Both elementary steps are good approximations if all other rates are small compared to $\Omega$. 

As a result of the SDRG procedure, the sites of the original lattice are organized into clusters which are characterized by effective annihilation rates and having negligible interactions with each other. 
The approach of the persistence probability by the SDRG method was described generally in Ref. \cite{jk2020}. 
It relies on that the renormalization procedure mimics the time evolution of the system and provides the set of sites occupied with an $O(1)$ probability at time $t$. The constituents of clusters that have been eliminated until the scale $\Omega=1/t$, will be unoccupied with a high probability at time $t$. 
The constituents of those clusters which have not been eliminated yet at scale $\Omega=1/t$ are occupied with a high probability if and only if any of their constituent sites was occupied initially (at time $t=0$). 

As mentioned in the previous section, the relevant interactions taken into account by the SDRG approach at any stage of the procedure are the nearest-neighbor ones, hence it is enough to consider the persistence of the first site of a semi-infinite chain. Therefore we shall recapitulate the way of calculating the persistence for this particular case.  
We make use of the representation of $P_0(t)$ as a return probability, set $\mu_0=0$ and assume that initially only the first site (labeled by $0$) is occupied. Let us use the variable $p$ for characterizing the return probability to the initial state (which is equivalent with the persistence of site $0$). Initially $p=1$, and it remains unchanged until an interaction term between site $0$ and the neighboring cluster (labeled by $n$) is picked for decimation, i.e. $\Omega=\lambda_{0n}$. The newly formed cluster which is composed of site $0$ and cluster $n$ has a simple internal dynamics: site $0$ is always occupied (since $\mu_0=0$), while the other constituent becomes occupied with rate $\lambda_{0n}$ and unoccupied with rate $\mu_n$. In the steady state of the new cluster, the probability that cluster $n$ is unoccupied (which is the return probability to the initial state) is simply $\mu_n/(\mu_n+\lambda_{0n})$.
Thus, upon decimating the interaction term $\lambda_{0n}$ of site $0$, or, at a later stage that of the cluster containing site $0$, the variable $p$ transforms as 
\be
\tilde p=p\frac{\mu_n}{\Omega+\mu_n}.
\label{p}
\ee

\section{Persistence by the SDRG method in the PL model}
\label{sec:pl}

\subsection{SDRG scheme of the PL model}

An analytically tractable SDRG scheme for the one-dimensional PL model has been formulated in Ref. \cite{qlrcp}. It is valid in the inactive phase and in the critical point, where decimations of interaction terms are only a vanishing fraction of total decimations at low rate scales. As a consequence, the spatial extension of clusters is much smaller than the spacings between them. In the simplified scheme, called as primary scheme in Ref. \cite{qlrcp}, the interactions between clusters are approximated by the long-range interactions between the closest constituent sites. Since only the interaction terms between neighboring clusters are renormalized at any stage, the procedure has, in effect, a one-dimensional structure, operating on the sequence of parameters, $\{\mu_n,\lambda_n\}$, where $\lambda_n$ denotes the nearest-neighbor creation rates.  
In terms of the reduced variables 
\be
\zeta=\left(\frac{\Omega}{\lambda}\right)^{1/\alpha}-1, \qquad 
\beta=\frac{1}{\alpha}\ln\frac{\Omega}{\mu},
\ee
the transformation rules read as 
\beqn
\tilde\beta=\beta_n+\beta_{n+1}-B  \qquad (\Omega=\lambda_n) \label{beta} \\
\tilde\zeta=\zeta_{n-1,n}+\zeta_{n,n+1}, \qquad (\Omega=\mu_n) 
\eeqn 
where $B=\frac{1}{\alpha}\ln 2$.

Since the rates $\mu_n$ and $\lambda_n$ remain independent during the SDRG procedure, it is sufficient to keep track of the evolution of their distributions, 
 $g_{\Gamma}(\beta)$ and $f_{\Gamma}(\zeta)$. 
Here, $\Gamma$ denotes a logarithmic renormalization scale 
\be
\Gamma\equiv\frac{1}{\alpha}\ln\frac{\Omega_0}{\Omega},
\label{Gamma}
\ee
where $\Omega_0$ is the initial value of $\Omega$. 

As can be seen in Eq. (\ref{beta}), which is equivalent to Eq. (\ref{mu}), the generated annihilation rate $\tilde\mu$ can happen to be greater than $\Omega$, which amounts to $\tilde\beta<0$, thus the rate scale does not decrease monotonically by such decimations. In this case, the formation of a cluster with $\tilde\mu>\Omega$ is immediately followed by its elimination. The composition of these two subsequent steps can be regarded as a single (anomalous) renormalization step, restoring thereby formally the monotonicity of $\Omega$.  
In the analytic SDRG description presented in Ref. \cite{qlrcp}, such anomalous decimations were not taken into account in their full complexity. In spite of this, the main characteristics of the fixed-point solutions could be determined, but the distribution $g_{\Gamma}(\beta)$ remained undetermined. 
It turns out, however, that taking the anomalous decimations into account properly, allows us to obtain both $f_{\Gamma}(\zeta)$ and $g_{\Gamma}(\beta)$.    
Under the repeated application of the renormalization rules specified above, we find that the distributions evolve according to the master equations 
\beqn  
\frac{\partial g_{\Gamma}(\beta)}{\partial\Gamma}=
\frac{\partial g_{\Gamma}(\beta)}{\partial\beta} +  \nonumber \\
+f_0(\Gamma)\int_0^{\beta+B} d\beta'g_{\Gamma}(\beta')g_{\Gamma}(\beta-\beta'+B) +
g_{\Gamma}(\beta)[g_0(\Gamma)-f_0(\Gamma)s(\Gamma)],  \label{gflow} \\
\frac{\partial f_{\Gamma}(\zeta)}{\partial\Gamma}=
(\zeta+1)\frac{\partial f_{\Gamma}(\zeta)}{\partial\zeta} + \nonumber \\  
+\{g_0(\Gamma)+f_0(\Gamma)[1-s(\Gamma)]\}
\int_0^{\zeta} d\zeta'f_{\Gamma}(\zeta')f_{\Gamma}(\zeta-\zeta') +
f_{\Gamma}(\zeta)[f_0(\Gamma)s(\Gamma)+1-g_0(\Gamma)], \nonumber \\
\label{fflow}
\eeqn
where $g_0(\Gamma)\equiv g_{\Gamma}(0)$, $f_0(\Gamma)\equiv f_{\Gamma}(0)$, and 
$s(\Gamma)=\int_0^{\infty}d\beta_1\int_0^{\infty}d\beta_2g_{\Gamma}(\beta_1)g_{\Gamma}(\beta_2)\Theta(\beta_1+\beta_2-B)$ is the probability that $\tilde\beta>0$. Here, $\Theta(x)$ denotes the Heaviside step function. 
These equations can be solved by the ansatz
\beqn
f_{\Gamma}(\zeta)=f_0(\Gamma)e^{-f_0(\Gamma)\zeta} \label{fp_f} \\
g_{\Gamma}(\beta)=g_0(\Gamma)e^{-g_0(\Gamma)\beta},  \label{fp_g}
\eeqn
which leads to the flow equations
\beqn
\frac{df_0}{d\Gamma}=f_0[1-g_0-f_0(1-s)], \\
\frac{dg_0}{d\Gamma}=-f_0g_0e^{-g_0B},
\eeqn
where $s=e^{-g_0B}(1+g_0B)$. 
The special case $B=0$ of these equations appear in the SDRG description of the random transverse-field Ising chain with long-range interactions \cite{jki}, and also in a random quantum rotor model \cite{akpr}. In Ref. \cite{akpr}, it was also numerically demonstrated that the solution in Eqs. (\ref{fp_f}-\ref{fp_g}) is a stable attractor for different forms of initial disorder distributions.    

Depending on the initial distribution of transition rates, the system can flow to different fixed points in the limit $\Gamma\to\infty$. The inactive phase is described by a line of fixed points, for which $g_0\to const=\alpha/z>1$ and $f_0\to 0$, where $z$ can be interpreted as a dynamical exponent, which depends on the initial distribution. For large $\Gamma$, the parameters approach to their limits exponentially in terms of $\Gamma$:
\be
g_0(\Gamma)-\frac{\alpha}{z}\sim f_0(\Gamma) \sim e^{-(\frac{\alpha}{z}-1)\Gamma}.
\label{f_a}
\ee     
In the critical fixed point, they tend to the limits $g_0\to 1$ and $f_0\to 0$, asymptotically as 
\beqn 
g_0(\Gamma)=1+\frac{2}{\Gamma}+O(\Gamma^{-2})  \\
f_0(\Gamma)=\frac{2e^B}{\Gamma^2}+O(\Gamma^{-3}).
\label{f_c}
\eeqn

\subsection{Renormalization of persistence}

We will now investigate what implications the SDRG approach has on the persistence. As said above, we consider the persistence of the first site ($0$) of a semi-infinite lattice. Here, the annihilation rate is set to zero ($\mu_0=0$), which guarantees that this site (or, at a later stage, the cluster containing this site) is never eliminated. Each time the creation rate between the surface cluster (containing site $0$) and the neighboring one is picked for decimation, the two clusters are merged and, at the same time, the variable $p$ is renormalized according to Eq. (\ref{p}).

Let us consider the number $n$ of such renormalization events, and calculate first the probability $Q_0({\Gamma})$ that no such event occurs ($n=0$) up to scale $\Gamma$.
Let $q_{\Gamma}(\zeta)$ denote the probability distribution of the first creation rate $\zeta$ under the condition that it has not been decimated yet. 
We find that its evolution is governed by  
\beqn
\frac{\partial q_{\Gamma}(\zeta)}{\partial\Gamma}=
(\zeta+1)\frac{\partial q_{\Gamma}(\zeta)}{\partial\zeta} + \nonumber \\  
+ [g_0+f_0(1-s)]
\int_0^{\zeta} d\zeta'q_{\Gamma}(\zeta')f_{\Gamma}(\zeta-\zeta') + \nonumber \\ 
+ q_{\Gamma}(\zeta)[q_0-f_0+f_0s+1-g_0],
\eeqn
where $q_0(\Gamma)=q_{\Gamma}(0)$. 
Comparing this equation with Eq. (\ref{fflow}), we obtain the result that $q_{\Gamma}(\zeta)\equiv f_{\Gamma}(\zeta)$. In words, the distribution of the creation rate of the first site under the condition that it has not yet been decimated is identical to the distribution of $\zeta$ in the bulk of the chain.   
By changing $\Gamma$ to $\Gamma+d\Gamma$, $Q_0({\Gamma})$ is thus reduced by 
$dQ_0=-Q_0f_0d\Gamma$, and we have
\be 
Q_0({\Gamma})=\exp\left[-\int_{\Gamma_0}^{\Gamma}f_0(\Gamma')d\Gamma'\right].
\ee
The mean number $\overline{n_{\Gamma}}$ of renormalization events can also be easily calculated as
\be  
\overline{n_{\Gamma}}=\int_{\Gamma_0}^{\Gamma}f_0(\Gamma')d\Gamma'.
\ee
In fact, since the distribution of the actually first creation rate depends only on $\Gamma$, the number of decimation events follows a Poisson distribution
\be 
Q_n({\Gamma})=e^{-\overline{n_{\Gamma}}}\frac{(\overline{n_{\Gamma}})^n}{n!}.
\label{poisson}
\ee

Using Eq. (\ref{f_a}), we obtain for the mean number of renormalization events
$\overline{n_{\Gamma}}\sim  e^{-(\frac{\alpha}{z}-1)\Gamma_0}- e^{-(\frac{\alpha}{z}-1)\Gamma}$ in the inactive phase, which converges in the limit $\Gamma\to\infty$. This is compatible with the expectation that the persistence probability tends to a non-zero limit here. Surprisingly, in the critical point, Eq. (\ref{f_c}) implies that $\overline{n_{\Gamma}}$ still converges to a finite constant, although more slowly than in the inactive phase: 
\be 
\overline{n_{\Gamma}}=2e^B(\Gamma_0^{-1}-\Gamma^{-1}).
\ee
This suggests that the persistence tends to a non-zero limit also in the critical point. 

After having determined the number of renormalization events we proceed with the distribution of the persistence. Introducing the variables
\be
K=\ln\frac{1}{p}, \qquad \gamma=\ln\frac{\Omega+\mu}{\mu}, 
\label{logpers}
\ee
the transformation rule in Eq. (\ref{p}) can be written as 
\be 
\tilde K = K + \gamma.
\label{Kgamma}
\ee
We can then formulate the following evolution equation for the distribution of $K$, $B_{\Gamma}(K)$: 
\be 
\frac{\partial B_{\Gamma}(K)}{\partial\Gamma}=-f_0\left[B_{\Gamma}(K)-\int_{0}^{K-\ln 2}B_{\Gamma}(K')h_{\Gamma}(K-K')\right],
\label{Keq}
\ee
where $h_{\Gamma}(\gamma)$ denotes the distribution of $\gamma$.
Using the fixed-point solution in Eq. (\ref{fp_g}), we have 
\be 
h_{\Gamma}(\gamma)=\frac{g_0}{\alpha}e^{\gamma}(e^{\gamma}-1)^{-1-g_0/\alpha}. 
\ee
It is not a simple exponential, furthermore, 
the positive lower bound $\ln 2$ of $\gamma$ implies that the solution must be nonanalytic at $K=n\ln 2$, $n=1,2,\cdots$. For these reasons, it is difficult to find the complete solution of Eq. (\ref{Keq}). Nevertheless, there are some general features of the solution which can be inferred from the form of Eq. (\ref{Keq}).
Since $g_0$ tends to a non-zero constant as $\Gamma\to\infty$, both in the inactive phase and in the critical point, the variable $\gamma$ has a limit distribution. We can see in Eq. (\ref{Keq}) that the derivative $\frac{\partial B_{\Gamma}(K)}{\partial\Gamma}$ is proportional to $f_0(\Gamma)$, therefore $B_{\Gamma}(K)$ must converge to a limit distribution as
$B_{\Gamma}(K)=B_{\infty}(K)+O[e^{-(\frac{\alpha}{z}-1)\Gamma}]$ in the inactive phase and as 
\be 
B_{\Gamma}(K)=B_{\infty}(K)+O(\Gamma^{-1})
\ee 
in the critical point. 
From Eq. (\ref{Keq}), the evolution equation of the average directly follows: 
\be 
\frac{d\overline{p_{\Gamma}}}{d\Gamma}=-f_0\overline{p_{\Gamma}}(1-\overline{e^{-\gamma}}).
\ee
For large $\Gamma$, it has a solution of the form 
$\overline{p_{\Gamma}}=\overline{p_{\infty}}+Ce^{-(\frac{\alpha}{z}-1)\Gamma}$ in the inactive phase and 
\be 
\overline{p_{\Gamma}}=\overline{p_{\infty}}+C\Gamma^{-1}
\ee 
in the critical point, where $C$ stands for a positive constant. 

The dependence of persistence on $\Gamma$ obtained by the primary SDRG scheme can be translated to time-dependence by the substitution $\Omega=1/t$. 
In the inactive phase, this yields an algebraic approach of the average persistence to a constant, $\overline{P_0(t)}-\overline{P_0(\infty)}\sim t^{\frac{1}{\alpha}-\frac{1}{z}}$, 
whereas, in the critical point, we obtain a logarithmically slow convergence,  
$\overline{P_0(t)}-\overline{P_0(\infty)}\sim [\ln(t/t_0]^{-1}$. 
The primary scheme is based on the assumption that the interaction between adjacent clusters is dominated by the long-range interaction between their closest constituents. This approach can be improved by taking into account the long-range interactions between all pairs of constituents of neighboring clusters. As it was argued in Refs. \cite{jki,qlrcp}, this leads to that, in the results obtained by the primary scheme, $\Omega$ must be replaced by $\Omega/m_{\Gamma}^2$, where the mean number $m_{\Gamma}$ of constituents of clusters is $m_{\Gamma}\sim \Gamma^2$ in the critical point \cite{qlrcp} and $m_{\Gamma}\sim \Gamma$ in the inactive phase \cite{star}.
Then the corrected time-dependence of the average persistence reads as 
\be 
\overline{P_0(t)}-\overline{P_0(\infty)}\sim [t\ln^2(t/t_0)]^{\frac{1}{\alpha}-\frac{1}{z}}
\label{Pt_i}
\ee
in the inactive phase and 
\be 
\overline{P_0(t)}-\overline{P_0(\infty)}\sim \{\ln[(t/t_0)\ln^4(t/t_0)]\}^{-1}
\label{Pt_c}
\ee
in the critical point. 

\section{Persistence by the SDRG method in the SE model}
\label{sec:se}

\subsection{SDRG scheme of the SE model}

The SDRG scheme of the SE model has a one-dimensional structure, similar to that of the PL model \cite{stretched}. The interactions between adjacent clusters are dominated by the long-range interaction between closest constituents. 
The reduced variables which transform additively are 
\be
\zeta=\left[\ln(\lambda_0/\lambda)/\Gamma\right]^{1/a}-1, \qquad 
\beta=\ln\frac{\Omega}{\mu},
\ee
where $\lambda$ denotes the effective creation rate between neighboring clusters and $\Gamma=\ln(\lambda_0/\Omega)$. 
The transformation rules look the same as those of the PL model,  
\beqn
\tilde\beta=\beta_n+\beta_{n+1}-B  \qquad (\Omega=\lambda_n)  \\
\tilde\zeta=\zeta_{n-1,n}+\zeta_{n,n+1}, \qquad (\Omega=\mu_n) 
\eeqn 
with $B=\ln 2$. The first one of these is valid both in the inactive phase and in the critical point, however, the second one is valid only in the inactive phase, since otherwise the width of clusters is not negligible compared to the distance between them \cite{stretched}.  
The distributions of reduced variables, $g_{\Gamma}(\beta)$ and $f_{\Gamma}(\zeta)$ evolve according to the master equations
\beqn  
\frac{\partial g_{\Gamma}(\beta)}{\partial\Gamma}=
\frac{\partial g_{\Gamma}(\beta)}{\partial\beta} +  \nonumber \\
+\frac{f_0(\Gamma)}{a\Gamma}\int_0^{\beta+B} d\beta'g_{\Gamma}(\beta')g_{\Gamma}(\beta-\beta'+B) +
g_{\Gamma}(\beta)[g_0(\Gamma)-\frac{f_0(\Gamma)}{a\Gamma}s(\Gamma)],  \label{gflow_se} \\
\frac{\partial f_{\Gamma}(\zeta)}{\partial\Gamma}=
\frac{\zeta+1}{a\Gamma}\frac{\partial f_{\Gamma}(\zeta)}{\partial\zeta} + \nonumber \\  
+\{g_0(\Gamma)+\frac{f_0(\Gamma)}{a\Gamma}[1-s(\Gamma)]\}
\int_0^{\zeta} d\zeta'f_{\Gamma}(\zeta')f_{\Gamma}(\zeta-\zeta') +
f_{\Gamma}(\zeta)[\frac{f_0(\Gamma)}{a\Gamma}s(\Gamma)+\frac{1}{a\Gamma}-g_0(\Gamma)], \nonumber \\
\label{fflow_se}
\eeqn
Again, Eq. (\ref{gflow_se}) is valid in the inactive phase and in the critical point, whereas  Eq. (\ref{fflow_se}) only in the inactive phase. 
The solutions are of the form given in Eqs. (\ref{fp_f}-\ref{fp_g}) with the parameters $f_0$ and $g_0$ obeying the flow equations 
\beqn
\frac{df_0}{d\Gamma}=f_0\left[\frac{1}{a\Gamma}-g_0-\frac{f_0}{a\Gamma}(1-s)\right], \\
\frac{dg_0}{d\Gamma}=-\frac{f_0}{a\Gamma}g_0e^{-g_0B},
\eeqn
where $s=e^{-g_0B}(1+g_0B)$. The special case $B=0$ of these equations, which describes a random transverse-field Ising chain with ferromagnetic, SE interactions, was derived in Ref. \cite{stretched}.

The inactive phase corresponds to a line of fixed points, at which $f_0$ tends to zero, whereas $g_0$ tends to a constant ($1/z>0$) depending on the initial disorder as  
\beqn 
f_0(\Gamma) \sim \Gamma^{1/a}e^{-\Gamma/z},  \\
g_0(\Gamma) - 1/z \sim  \Gamma^{1/a-1}e^{-\Gamma/z}.
\eeqn
In the critical point, where formally $1/z=0$, the fixed-point distribution $f_{\Gamma}(\zeta)$ is unknown, nevertheless it was shown in Ref. \cite{stretched} that 
\beqn 
f_0(\Gamma) \simeq a,  \\
g_0(\Gamma) \sim \frac{1-a}{a\Gamma}
\eeqn
for large $\Gamma$. 

\subsection{Renormalization of persistence}

As for the PL model, it is sufficient to consider the persistence of the first site of a semi-infinite system. 
We begin with the calculation of the distribution of renormalization events of persistence. 
Similar to the PL model, one can show that the distribution of the creation rate of the surface cluster is the same as in the bulk of the system. 
Consequently, when the renormalization scale is increased by $d\Gamma$, the 
probability of decimating the creation rate of the first cluster is $\frac{f_0}{a\Gamma}d\Gamma$, and we can write for the mean number of decimation events:
\be  
\overline{n_{\Gamma}}=\int_{\Gamma_0}^{\Gamma}\frac{f_0(\Gamma')}{a\Gamma'}d\Gamma'.
\ee
This converges in the inactive phase as $\overline{n_{\Gamma}}=\overline{n_{\infty}}+O(\Gamma^{1/a-1}e^{-\Gamma/z})$, indicating a non-zero average persistence at late times, while it diverges in the critical point as $\overline{n_{\Gamma}}=\ln(\Gamma/\Gamma_0)$ pointing toward a vanishing persistence in the limit $t\to\infty$.  
As for the PL model, the number $n$ of decimation events follows a Poisson distribution given in Eq. (\ref{poisson}).  

Using the logarithmic variables in Eq. (\ref{logpers}), 
the transformation of persistence becomes additive as given in Eq. (\ref{Kgamma}), and the distribution $B_{\Gamma}(K)$ obeys the master equation 
\be 
\frac{\partial B_{\Gamma}(K)}{\partial\Gamma}=-\frac{f_0}{a\Gamma}\left[B_{\Gamma}(K)-\int_{0}^{K-\ln 2}B_{\Gamma}(K')h_{\Gamma}(K-K')\right],
\label{master}
\ee
where $h_{\Gamma}(\gamma)$ denotes the distribution of $\gamma$.
In the inactive phase, the form of this equation implies that $B_{\Gamma}(K)$ converges to a limit distribution as 
$B_{\Gamma}(K)=B_{\infty}(K)+O(\Gamma^{1/a-1}e^{-\Gamma/z})$. 
This also implies that the average persistence tends to a constant as 
$\overline{p_{\Gamma}}=\overline{p_{\infty}}+O(\Gamma^{1/a-1}e^{-\Gamma/z})$.

In the critical point, we can make an approximation for large $\Gamma$, which greatly simplifies Eq. (\ref{master}). Since $g_0(\Gamma)\to 0$, typically $\beta\gg 1$ for large $\Gamma$, and we can write the transformation of persistence  as 
\be 
\tilde K=K+\gamma=K+\beta+\ln(1+e^{-\beta})\simeq K+\beta.
\ee 
Eq. (\ref{master}) then simplifies to
\be 
\frac{\partial B_{\Gamma}(K)}{\partial\Gamma}\simeq-\frac{f_0}{a\Gamma}\left[B_{\Gamma}(K)-\int_{0}^{K}B_{\Gamma}(K')g_{\Gamma}(K-K')\right].
\label{master2}
\ee
This has the fixed-point solution
\be 
B_{\Gamma}(K)\simeq g_0e^{-g_0K}, 
\ee
which results in 
\be 
\overline{p_{\Gamma}}\simeq \frac{g_0}{1+g_0} \sim \Gamma^{-1}
\ee
for the average persistence. 
The time-dependence of persistence can be obtained by the substitution $\Omega=1/t$, which results in 
$\overline{P_0(t)}-\overline{P_0(\infty)}\sim t^{-1/z}[\ln(t/t_0)]^{1/a-1}$
in the inactive phase. Similar to the PL model in the inactive phase, this result can be improved by taking into consideration the long-range interactions between all pairs of constituents of adjacent clusters, which can be achieved by the replacement $\Omega\to\Omega/\Gamma^2$, yielding  
\be
\overline{P_0(t)}-\overline{P_0(\infty)}\sim t^{-1/z}[\ln(t/t_0)]^{1/a-1-2/z}
\ee
In the critical point, the simple SDRG scheme which relies on the dominance of the long-range interaction of closest constituents of clusters results in   
\be
\overline{P_0(t)} \sim [\ln(t/t_0)]^{-1}.
\ee
Unlike in the active phase, the size of clusters is comparable with the spacing between, hence a similar {\it a posteriori} improvement of this scheme cannot be easily achieved. 
Nevertheless, long-range interactions between the bulk constituents of clusters must give corrections to this form.

\section{Effects of rare regions in the active phase}
\label{phenomenology}

It has been known for a long time that the temporal decay of the density in the inactive phase of the disordered contact process is anomalously slow, due to the occurrence of rare, locally supercritical domains, which have a long lifetime \cite{noest}. Similarly, in the active phase of the model the average persistence exhibits a slower-than-exponential decay due to the presence of locally subcritical regions, as it has been pointed out in the short-range model in Ref. \cite{jk2020} and observed also in a similar model in Ref. \cite{bg}. 

The simple phenomenological considerations based on the occurrence of rare, subcritical regions, which were formulated for the short-range model in $d$ dimensions \cite{jk2020}, can easily be generalized to the case of long-range interactions. 
The starting point is that roughly isotropic, compact, subcritical regions of radius $\ell$ can occur anywhere in the system with a probability 
\be 
P_>(\ell)\sim e^{-A\ell^d},
\ee  
where $A$ is a positive, non-universal constant. 
The central site of such a region, provided it was initially empty, will predominantly lose its persistence through a creation event from outside of the rare region. The total rate of this event is 
$\lambda_{\rm total}\sim \int_{\ell}^{\infty}\lambda(r)r^{d-1}dr$, and the corresponding time scale is   
$\tau \sim 1/\lambda_{\rm total}\sim \ell^{d-\alpha}$ in the PL model and  
$\tau \sim 1/\lambda_{\rm total}\sim e^{(\ell/l_0)^a}\ell^{a-d}$ in the SE model for $0<a<1$.
The average persistence at late times is then given by
\be 
\overline{P(t)} \sim \int_{\ell_0}^{\infty}e^{-t/\tau(\ell)}\rho(\ell)d\ell, 
\ee 
where $\rho(\ell)\sim e^{-A\ell^d}\ell^{d-1}$ is the probability density of the radius of rare regions. 
For the PL model, one finds that the saddle point of the integrand is at $\ell^*\sim t^{1/\alpha}$, yielding
\be 
\overline{P(t)} \sim \exp\{-Ct^{d/\alpha}+O(\ln t)\},       \qquad (PL)
\label{Pt_a}
\ee
where $C$ denotes a positive, non-universal constant. 
A similar calculation for the SE model gives $\ell^*\sim [\ln(t/t_0)]^{1/a}$, which results in 
\be 
\ln[\overline{P(t)}] \sim -C[\ln(t/t_0)]^{d/a} \qquad (SE)
\ee  
in leading order. Here, $C$ and $t_0$ denote positive, non-universal constants again.

\section{Numerical results}
\label{numerical}

In order to check the predictions of the SDRG method and the phenomenological rare-region arguments, we performed Monte Carlo simulations and calculated the time-dependence of the average persistence. As the SDRG results for the SE model are similar to those of the short-range model \cite{jk2020}, we concentrated on the PL model, which represents a frequently studied, ubiquitous form of interactions, and for which the SDRG predictions are qualitatively different from those of the short-range model.  Another advantage of the PL model is that estimates of the critical point are available from Ref. \cite{qlrcp}. 

We considered a dilution type of disorder, i.e. the sites of the lattice are removed randomly with a probability $1/2$. The simulations were implemented as follows. An occupied site is randomly selected and made unoccupied with a probability $1/(1+\lambda_0)$, or, with a probability $\lambda_0/(1+\lambda_0)$ the creation of a new particle is attempted. To select a target site, a random variable  $1<r<\infty$ is generated from the distribution $\rho(r)=(\alpha-1)r^{-\alpha}$ and the integer part of $r~{\rm mod}~L$, where $L$ is the size of the system, is calculated. Fixing in this way the distance of the target site from the source, one of the two candidates is picked with equal probabilities. If the target site is an existing, empty site, it is made occupied. 
The time increment associated with such an update step is $\Delta t=1/N(t)$, where $N(t)$ is the number of occupied sites. 
The system size was $L=10^9$ in the active phase and in the critical point, and $L=10^8$ in the active phase, where simulations are slower owing to the non-vanishing density of particles. 
We started simulations from an initial state in which the sites were occupied randomly with a probability $1/2$, and measured the fraction of persistent sites as a function of time. An average over data obtained in $10-100$ different random realizations of disorder was also performed.           
We present results obtained with $\alpha=2$, for which the critical point was estimated to be at $\lambda_0=2.90(1)$ and for which the predictions of the SDRG method on the time-dependence of the order parameter have been confirmed by simulations \cite{qlrcp}. 

\begin{figure}[h]
\includegraphics[width=0.7\linewidth]{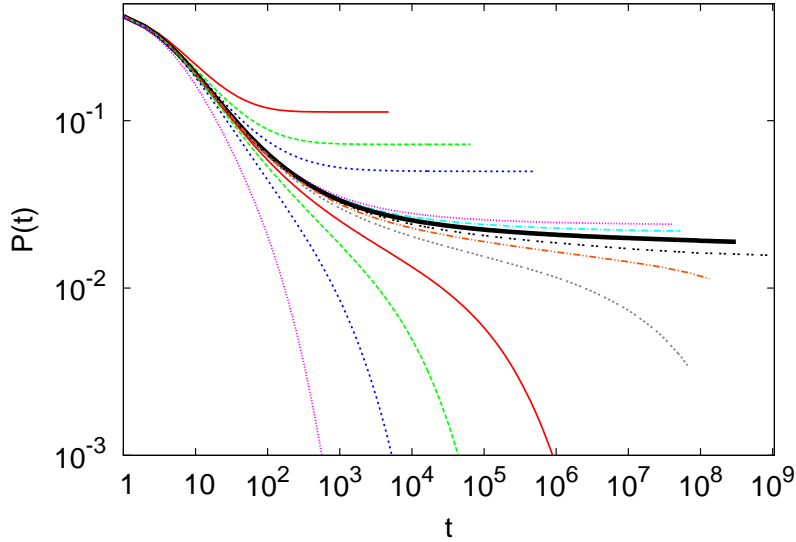}
\caption{
Time-dependence of the average persistence obtained by numerical simulations for the PL model with $\alpha=2$, for different values of the control parameter 
$\lambda_0=2.5,2.7,2.8,2.89,2.895,2.9,2.905,2.91,2.92,2.95,3,3.1,$ and $3.5$ (from top to bottom). The critical curve at $\lambda_0=2.9$ is plotted by a thick black line. 
\label{p2c.5}
}
\end{figure}

The average persistence probability as a function of time is shown in Fig. \ref{p2c.5} for different values of the control parameter $\lambda_0$.
As can be seen, in the active phase ($\lambda_0>\lambda_c=2.90$), $\overline{P(t)}$ tends to a non-zero limit at late times, and this seems to hold also for the critical curve, although the convergence is slower.  Deeply in the inactive phase, $\overline{P(t)}$ decreases rapidly to zero, and approaching the critical point, the cutoff is shifted to later and later times.  

\begin{figure}[ht]
\includegraphics[width=0.7\linewidth]{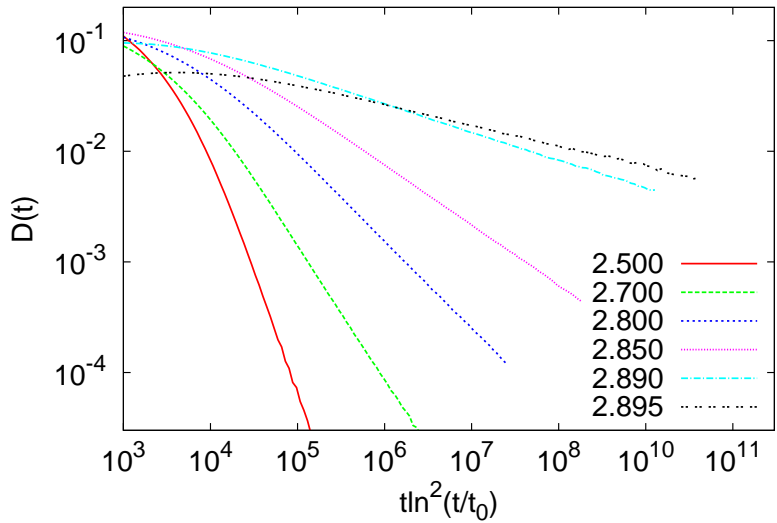}
\caption{
Time-dependence of the derivative $D(t)$ of the average persistence defined in Eq. (\ref{Ddef}) for different values of the control parameter $\lambda_0$ in the inactive phase. The times scales are, in order, $t_0=1,10,20,30,100$, and $500$ for increasing $\lambda_0$. According to Eq. (\ref{Dt}), the curves must be linear  in this plot, with a slope $\frac{1}{\alpha}-\frac{1}{z}$.   
\label{p2c.5tp}
}
\end{figure}

First, let us have a closer look at the inactive phase. According to the SDRG method, the finite-time deviation of the average persistence from its limiting value is algebraic in time with a logarithmic correction, as given in Eq. (\ref{Pt_i}). To get rid of the unknown constant $\overline{P_0(\infty)}$ in Eq. (\ref{Pt_i}), we consider the derivative 
\be
D(t)=\frac{d\ln\overline{P(t)}}{d\ln[t\ln^2(t/t_0)]}=
\frac{d\ln\overline{P(t)}}{d\ln t}\left[1+\frac{2}{\ln(t/t_0)}\right]^{-1},
\label{Ddef}
\ee
which must tend to zero as 
\be
D(t) \sim [t\ln^2(t/t_0)]^{\frac{1}{\alpha}-\frac{1}{z}}.
\label{Dt} 
\ee   
for large $t$. 
As it is shown in Fig. \ref{p2c.5tp}, the time-dependence of the derivative indeed follows the law in Eq. (\ref{Dt}) with an exponent $\frac{1}{\alpha}-\frac{1}{z}$ varying with the control parameter.  

\begin{figure}[h]
\includegraphics[width=0.7\linewidth]{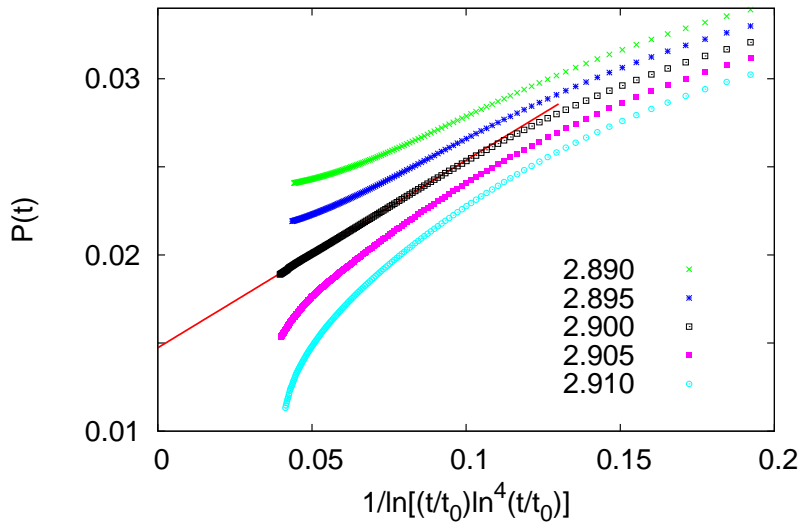}
\caption{
Time-dependence of the average persistence for different values of the control parameter $\lambda_0$ in the neighborhood of the critical point ($\lambda_c=2.90$). The times scale $t_0=150$ was used. According to Eq. (\ref{Pt_c}), the critical curve must be asymptotically linear in this plot. The solid line is a linear fit to the data for $\lambda_0=2.90$.  
\label{pfig}
}
\end{figure}

Next, let us consider the average persistence in the critical point. According to the results of the SDRG approach, it must tend to a non-zero limit logarithmically slowly in time, as given in Eq. (\ref{Pt_c}). 
As can be seen in Fig. \ref{pfig}, the numerical results are compatible with this result, although the limiting value of the persistence is rather small.  

\begin{figure}[h]
\includegraphics[width=0.7\linewidth]{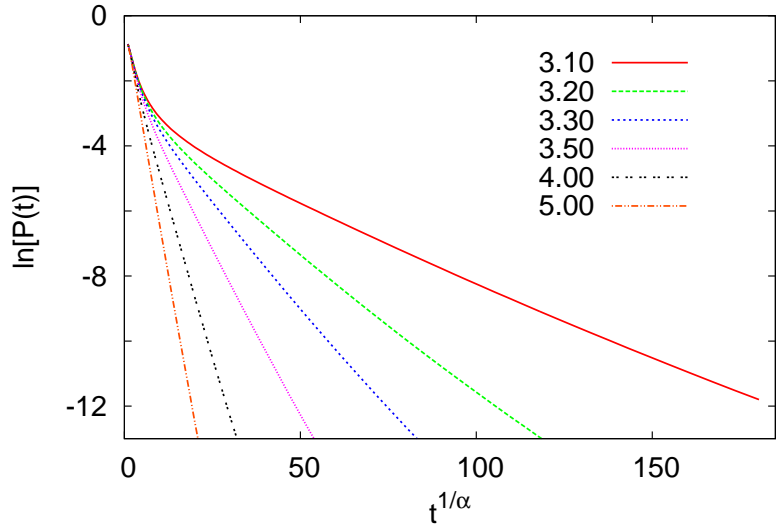}
\caption{
Time-dependence of the average persistence for different values of the control parameter $\lambda_0$ in the active phase of the PL model with $\alpha=2$.
\label{gp}
}
\end{figure}

The time-dependence of the average persistence in the inactive phase is shown in Fig. \ref{gp}. According to the result of the simple phenomenological considerations in Eq. (\ref{Pt_a}), the logarithm of the average persistence must be proportional to $t^{1/\alpha}$ in leading order. Deeply in the active phase, the numerical results seem to be compatible with this, but closer to the critical point, the curves are not straight asymptotically as they should be in the plot of Fig. \ref{gp}. This may be attributed to that the $O(\ln t)$ and other possible corrections are stronger closer to the critical point, and the leading term does not prevails at the times available by the simulations. Indeed, appropriately chosen logarithmic corrections can resolve this discrepancy (not shown), but the precise form of corrections cannot be determined with certainty from the present numerical data.    

\section{Discussion}
\label{discussion}

In this work, we studied the time-dependence of the local persistence during non-stationary time evolutions of the disordered contact process with long-range interactions by combining the SDRG method, phenomenological considerations and numerical simulations.  
For the PL model, the critical point is described by a finite-disorder fixed point of the SDRG transformation \cite{qlrcp}, at which the distribution of one of the variables (the reduced annihilation rates) does not broaden unboundedly, therefore the asymptotic exactness of the method is not guaranteed. In spite of this, the results of the method concerning the time-dependence of the order parameter for $\alpha>3/2$, where the Harris criterion predicts the relevance of disorder (and even for $\alpha=3/2$ for strong enough dilution), were found to be compatible with results of numerical simulations \cite{qlrcp}. For the validity of the approach, see also the argumentations in Ref. \cite{akpr}.  
We found by this method that the average persistence tends to a non-zero limit as $t\to\infty$, not only in the inactive phase but also in the critical point.
As the persistence tends to zero in the active phase, this means that the limiting value of the persistence probability is a discontinuous function of the control parameter.  

The possibility of such a phenomenon is closely related to the presence of long-range interactions. In critical, disordered contact processes the activity is concentrated on a set of clusters of occupied sites which comprise a vanishing fraction of the total system in the limit $t\to\infty$. In the short-range model, the interaction between constituents of a given cluster, which may be far from each other, take place through chains of creation events from one part of the cluster to another one along some path of sites external to the cluster. Any site which takes part in the mediation of the interaction by becoming part of such a path will loose its persistence. 
In a long-range model, however, the coherence of the different parts of cluster is predominantly realized by the existing long-range interactions, which does not risk the persistence of surrounding sites.    
Furthermore, in the PL model, the clusters of occupied sites are extremely sparse in the critical point (having a formally zero fractal dimension) compared to the short-range and SE models \cite{jki}, allowing for a macroscopic number of sites to remain intact.

Although the persistence tends to non-zero limits in the inactive phase and in the critical point alike, the form of the finite-time corrections reflects whether the system is critical or not. In the former case, the corrections vanish algebraically with exponents varying with the control parameter, while in the critical point, it decreases inversely proportionally to $\ln t$. This behavior of the correction term differs from that of the order parameter (the average density of occupied sites), which decays algebraically both in the inactive phase and in the critical point \cite{qlrcp}.   

In the SE model, the critical behavior is described by an infinite-disorder fixed point of the SDRG transformation, which ensures the validity of the approach. The critical behavior of this model is qualitatively similar to that of the short-range model, the difference appearing only in the critical exponents. We found that, as opposed to the PL model, the presence of stretched exponential interactions is not able to save a finite fraction of sites from losing their persistence in the critical point. According to our results, although the distribution of persistence is different from that of the short-range model, the average is found to decrease inversely proportionally to $\ln t$ just as in the short-range model \cite{jk2020}.  

We considered in this work one-dimensional models, but in the presence of long-range interactions, the spatial dimension is less important. As it was argued in Ref. \cite{qlrcp} for the PL model, the critical behavior of a $d$-dimensional model with a decay exponent $\alpha$ in the interaction strength is expected to be the same as that of a one-dimensional PL model with a reduced decay exponent $\alpha/d$. This was confirmed in dimensions $d=2$ \cite{qlrcp} and $d=3$ \cite{kji_3d} by Monte Carlo simulation and the numerically implemented SDRG method.   
Based on this, we expect the discontinuity of the persistence found in the $d=1$ PL model to appear also in higher dimensions and other values of $\alpha$, whenever the dimension is below the upper critical dimension $d_c=\min\{4,2(\alpha-d)\}$ \cite{janssen}, where the Harris criterion predicts weak disorder to be relevant \cite{qlrcp}.

\ack
The author thanks G. \'Odor for useful discussions.
This work was supported by the National Research, Development and Innovation Office -- NKFIH under grant No. K128989.

\end{document}